\documentclass[useAMS,a4paper,fleqn,usenatbib]{mnras}

\usepackage[T1]{fontenc}
\usepackage{ae,aecompl}
\usepackage{hyperref}
\usepackage{pdflscape}

\usepackage{graphicx}	
\usepackage{amsmath}	
\usepackage{amssymb}	
\usepackage{tabularx}
\usepackage{txfonts}
\usepackage{wasysym}
\usepackage{ragged2e}
\usepackage{natbib}
\usepackage[english]{babel}
\usepackage{float}
\usepackage{color}

\defcitealias{2016MNRAS.456.1137L}{Paper\,I}
\defcitealias{2016MNRAS.tmp.1056L}{Paper\,II}

\title[A PSF-based approach to \textit{Kepler/K2} data -- III]{A
  PSF-based approach to \textit{Kepler/K2} data -- III. Search for
  exoplanets and variable stars within the open cluster M\,67
  (NGC\,2682) \thanks{ Based on observations with the {\it Kepler}
    telescope and with the Schmidt 67/92\,cm Telescope at the
    Osservatorio Astronomico di Asiago, wich is part of the
    Osservatorio Astronomico di Padova, Istituto Nazionale di
    Astrofisica.}  }

\author[Nardiello et al.]{D.\ Nardiello$^{1,2}$\thanks{E-mail: domenico.nardiello@unipd.it}, 
M.\ Libralato$^{1,2}$, 
L.\ R.\ Bedin$^{2}$,
G.\ Piotto$^{1,2}$,
L.\ Borsato$^{1,2}$,
\newauthor
V. Granata$^{1,2}$,
L.\ Malavolta$^{1,2}$,
V.\ Nascimbeni$^{1,2}$ \\
$^{1}$Dipartimento di Fisica e Astronomia ``Galileo Galilei'', Universit\`a di Padova, Vicolo dell'Osservatorio 3, Padova IT-35122 \\
$^{2}$Istituto Nazionale di Astrofisica - Osservatorio Astronomico di Padova, Vicolo dell'Osservatorio 5, Padova, IT-35122 \\
}

\date{Accepted 2016 August 24. Received 2016 July 18; in original form 2016 July 18}

\pubyear{2016}

\begin{document}
\label{firstpage}
\pagerange{\pageref{firstpage}--\pageref{lastpage}}
\maketitle

\begin{abstract}
In the third paper of this series we continue the exploitation of {\it
  Kepler/K2} data in dense stellar fields using our PSF-based method.
This work is focused on a $\sim 720$-arcmin$^2$ region centred on
the Solar-metallicity and Solar-age open cluster M\,67.
We extracted light curves for all detectable sources in the {\it
  Kepler} channels 13 and 14, adopting our technique based on the usage
of a high-angular-resolution input catalogue and target-neighbour
subtraction.
We detrended light curves for systematic errors, and searched for
variables and exoplanets using several tools.
We found 451 variables, of which 299 are new detection.
Three planetary candidates were detected by our pipeline in this
field. 

Raw and detrended light curves, catalogues, and {\it K2} stacked
images used in this work will be released to the community.

\end{abstract}

\begin{keywords}
techniques: image processing -- techniques: photometric -- binaries:
general -- stars: variables: general -- exoplanets --
open clusters and associations: individual: M\,67
\end{keywords}



\section{Introduction} \label{intro}

The data collected during the reinvented {\it Kepler/K2} mission
(\citealt{2014PASP..126..398H}) allowed the community to search for new variable
stars and exoplanets in many Galactic fields, containing various kinds
of objects (single stars, open and globular clusters, extra-galactic
sources, etc.).  Despite the lower quality of {\it K2} data compared
to {\it Kepler} main mission, many techniques for the extraction and
the systematic correction of the light curves (LCs) have been developed in these last
two years.

\citet[hereafter \citetalias{2016MNRAS.456.1137L}]{2016MNRAS.456.1137L} developed a new tool
to extract high photometric precision LCs from the {\it K2} undersampled images
of crowded environments, based on the usage of effective Point Spread
Functions (ePSFs) and of a high-angular-resolution input catalogue.
However this approach is perfectly suitable for any stellar field.
This PSF-based technique also allows us to extract LCs for sources in
the faint magnitude regime ($K_{\rm P}>15.5$), increasing the number
of analysable objects in a field.

\begin{figure*}
\includegraphics[width=\textwidth]{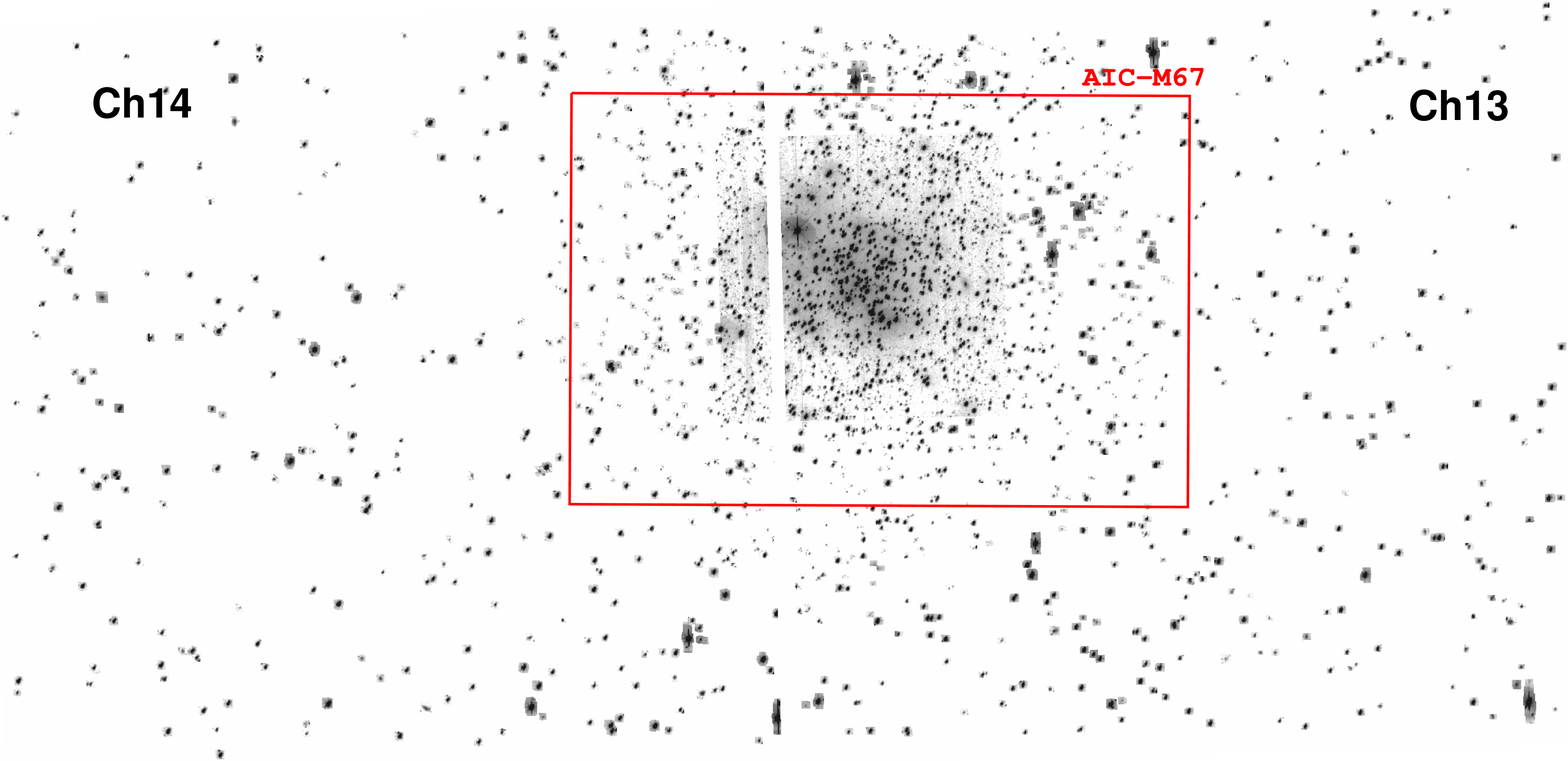}
\caption{Field of view covered by all available {\it K2}/C5 Ch13
  (right) and Ch14 (left) exposures used in our analysis. The red
  rectangle represents the field of view covered by the AIC-M67
  (\citealt{2016MNRAS.455.2337N}). The image is in logarithmic grey
  scale. North is up and East to the left. \label{fig:1}}
\end{figure*}

In this work we take advantage of this method, focusing our attention
on the moderate-crowded region containing the open cluster (OC) M\,67
(NGC\,2682). During the {\it K2} Campaign 5 ({\it K2}/C5), two
super-stamps (covering a region between two {\it Kepler} channels of
module 6) centred on M\,67 were achieved. We 
reconstructed all the images containing the super-stamps and applied
our PSF-based approach to extract high-precision LCs.

The OC M\,67 is one of the most studied and intriguing OCs in
literature (see, e.g., \citealt{2016MNRAS.455.2337N} and references
therein). This OC has an age and metallicity similar to that of the
Sun (e.g., \citealt{2010A&A...513A..50B, 2014A&A...561A..93H}) and is
located at a distance $<1$\,kpc (\citealt{2010MNRAS.403.1491P}).  In a
previous paper based on ground-based, Asiago Schmidt data
(\citealt{2016MNRAS.455.2337N}), we have already investigated this OC,
detecting 43 new variables. In this work, we used the same input
catalogue to extract LCs from the {\it K2}/C5 images, and search for
new variable stars among M\,67 members.

Together with M\,44
(\citealt{2012ApJ...756L..33Q,2016A&A...588A.118M})
and the Hyades (\citealt{2014ApJ...787...27Q}), M\,67 is one of the
few OCs that host stars with confirmed exoplanets.
\citet{2014A&A...561L...9B,2016A&A...592L...1B}, using radial velocity
(RV) measurements, have detected four exoplanets orbiting M\,67
members, three of which are main sequence (MS) stars. In this work we
conducted a search for transiting planets on all M\,67 member (and
not) stars, in order to identify low-mass planets that could have been
overlooked by RV searches.

\section{Observations and data reduction} \label{obs}

The {\it K2} observations were performed during C5. The data-set
includes 3620 usable long-cadence observations, that spanned over
74.82 days (April 27th - July 10th, 2015).

The bulk of M\,67 falls in between the two KEPLERCAM channels 13 (Ch13)
and 14 (Ch14) of module 6.  In this work we focus exclusively on the
point sources monitored by {\it K2}/C5 on these two channels.

In each channel a super-stamp was monitored.  The one in Ch14 is
90$\times$400 pixels$^2$, while the one of Ch13 is larger, with
312$\times$400 pixels$^2$. Overall the two super-stamps would
continuously cover a region of $402 \times 400$ pixels$^2$ ($\sim
0.2$\,degree$^2$) centred on the M\,67's centre (see Fig.~\ref{fig:1})
without considering the gap between the two channels.  We have
analysed as well the individual postage stamps present in both
channels.

We downloaded the {\it K2} Target Pixel Files (TPFs) containing the
complete time series data from the ``Mikulski Archive for Space
Telescopes''\,(MAST)\footnote{https://archive.stsci.edu/k2/data\_search/search.php}.
We reconstructed the 3620 images ($1132 \times 1070$ pixel$^2$) of the
series for each channel, one for each cadence number of the TPFs.  We
assigned to each pixel the sum of the values of the columns
\texttt{FLUX} and \texttt{FLUX\_BKG}. To each image we assigned the
average {\it Kepler} Barycentric Julian Day (KBJD) corresponding to
the KBJD associated to the cadence number.

In the following we give a brief summary of the LC extraction, that
was carefully described in \citetalias{2016MNRAS.456.1137L}. 
Our  method for the LC extraction essentially relies on: 
\begin{enumerate}
\item time-perturbed effective PSFs (ePSFs), 
\item a high-angular-resolution stellar input catalogue, 
\item six-parameters, local linear transformations between single-image catalogues and the input catalogue,
\item neighbour subtraction.
\end{enumerate}

\subsection{Improved PSFs}
In \citetalias{2016MNRAS.456.1137L} we have described how to model the undersampled PSF of
{\it Kepler} using the recipe by \citet{2000PASP..112.1360A}.
In this work we used a slightly improved version of the ePSFs obtained
with the method described in \citetalias{2016MNRAS.456.1137L}.
The method will be explained in detail in a future paper of the series
focus on M\,4 (Libralato et al. in preparation). It differs from that
presented in \citetalias{2016MNRAS.456.1137L} only in the
fact that neighbours are iteratively subtracted (using the current
ePSF model and an input list) to the sample stars used to define the
ePSF model and imposing the positions of the used stars from the input
list before computing the new improved ePSF models.

\subsection{The Asiago Input Catalogue for M\,67 (AIC-M67)}
We used as input catalogue the Asiago Schmidt catalogue of M\,67 released by
\citet{2016MNRAS.455.2337N}, hereafter
AIC-M67\footnote{http://groups.dfa.unipd.it/ESPG/VAR/M67/m67.dat}.
This catalogue, obtained using Asiago Schmidt 67/92\,cm data, gives
positions, magnitudes in 8 filters, proper motions and membership
probabilities for 6905 sources in a field of $58 \times
38$\,arcmin$^2$ centred on M\,67 cluster centre. The coverage of this
input catalogue is shown in Fig.~\ref{fig:1} (red rectangle).

\subsection{The {\it K2}-Stacked-Image Catalogues (K2S-Ch13/-Ch14)}
\label{aic}
Many M\,67 stars are located in postage stamps recorded by {\it K2}/C5 outside the AIC-M67 region in both channels (see Fig.~\ref{fig:1}), hence the need of extending the AIC-M67.
For each channel we made a stacked image using all 3620 usable
images, as described in \citetalias{2016MNRAS.456.1137L}.
Then, we applied the same procedure adopted on single images to 
extract the catalogues from the stacked images,  excluding all
 sources that fell inside the AIC-M67 region.
For the Ch13 and Ch14, we created two additional input catalogues: the
{\it K2}'s Stacked-Image Ch13 (hereafter K2S-Ch13) containing 437
additional stars, and the K2S-Ch14 with 328 additional stars.  For each
channel, the final input list adopted was a merge between the AIC-M67
and K2S-Ch13 or K2S-Ch14 catalogue.

\section{Light Curve extraction}

For the extraction of the LCs, we used the software developed and
described by \citet{2015MNRAS.447.3536N} for the ground-based
telescope Asiago Schmidt 67/92\,cm and adapted in \citetalias{2016MNRAS.456.1137L} for {\it
  K2} data.

Given a target star in the input catalogue, the software locally
transforms positions and magnitudes of all its input catalogue
neighbours (inside a radius of 35 {\it Kepler} pixels, i.e., $\sim
2.3$ arcmin) from the reference system of the input catalogue into that of the individual {\it K2} image.
Next, we subtracted these target neighbours from the considered image.
We extracted the flux of the target source from the original and the
neighbour-subtracted images, using three methods: (i) PSF-fitting,
(ii) aperture, and (iii) optimal-mask photometry.  
For aperture photometry we used four different apertures: 1-, 2.5-, 3.5-,
and 4.5- pixel radii.  For optimal-mask photometry, in analogy with
\citet{2014PASP..126..948V}, we used two different masks based on the
ePSF model: for the first mask (mask-1), suitable for bright stars
($K_{\rm P} <13$), are considered only pixels for which the
normalised ePSF value is $\ge 0.005$\%; the second mask (mask-2), suitable for
fainter stars, is made by pixels for which the ePSF value is $\ge 0.1$\%.

We have already demonstrated in \citetalias{2016MNRAS.456.1137L} that the
photometric precision is better for neighbour-subtracted photometries;
for this reason, hereafter, we will only consider neighbour-subtracted
LCs.

\subsection{Light Curve detrending} \label{lcproc}

The larger jitter of the spacecraft pointing during the {\it
  K2}-mission (if compared to that of the {\it Kepler} main mission) translates
into a worse photometric precision.
To correct the most of the systematic errors, that are related to the
spacecraft drift, several methods appeared in the literature.
In this work we used the same detrending algorithm presented in
\citetalias{2016MNRAS.456.1137L} for the {\it K2}/C0 Ch81 data, slightly improved
adding a new step, that proved to be effective in taking into account
the specifics of each campaign.

Already in the original {\it Kepler} mission the LCs show some
systematic effects not correlated with the positions and the magnitude
of the stars on the detectors, but associated with spacecraft,
detector and environment (\citealt{2013kepler.book.....R}). During a
{\it K2} campaign, all stars in the same channel show common
systematic trends in their LCs. This particular behaviour allows to model these
systematic trends as a linear combination of orthonormal functions,
called cotrending basis vectors (CBVs).  In an analogous way as the
{\it Kepler} standard pipeline for the LC reduction, we used the
publicly available\footnote{https://archive.stsci.edu/k2/cbv.html} CBVs (released by the {\it Kepler}
team) for modelling
and correcting these systematic features.

\begin{figure*}
\includegraphics[bb= 24 190 567 702, width=0.99\textwidth]{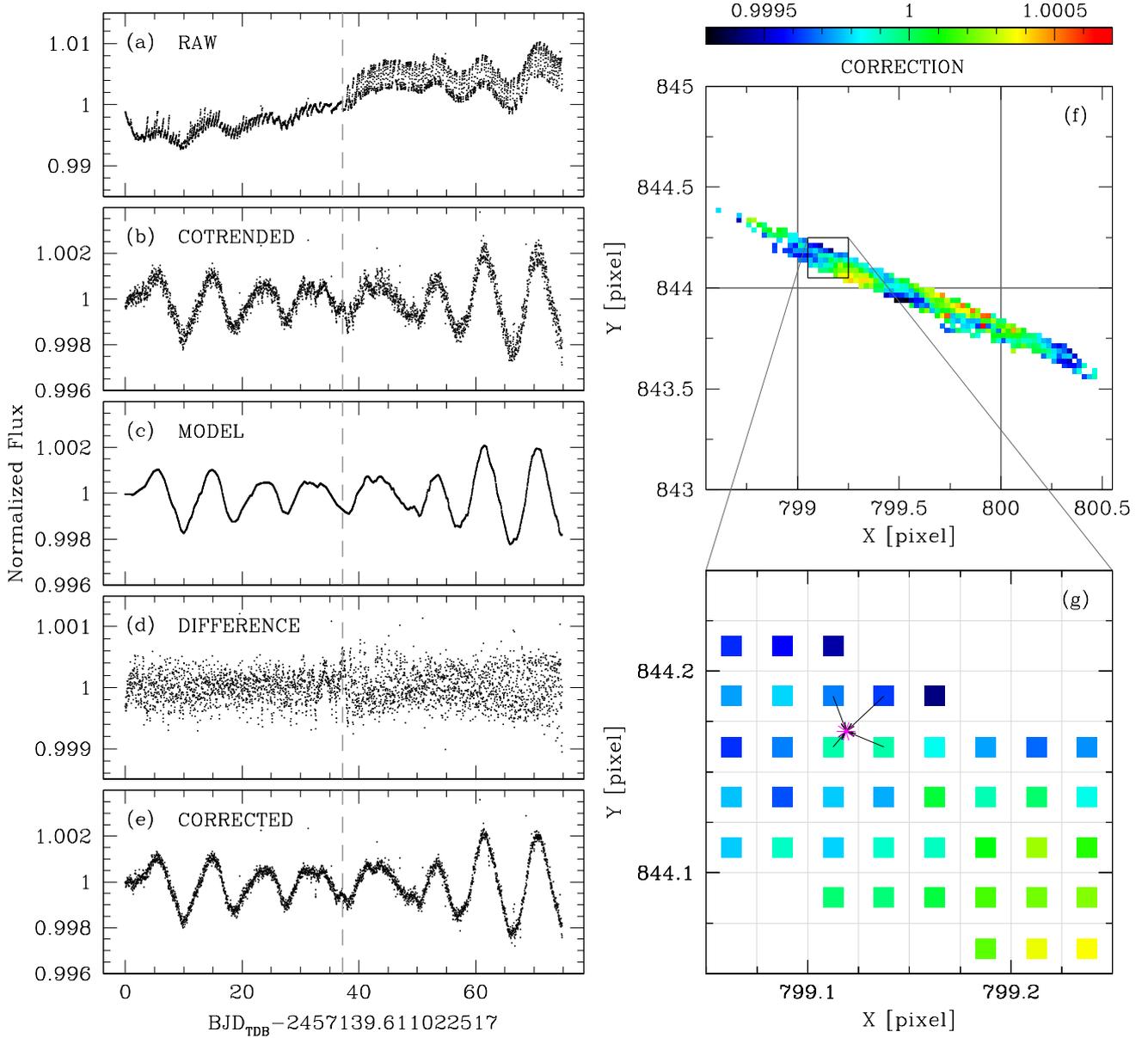}
\caption{Overview of the procedure used for correcting the
  3.5-pixel-aperture LC of star Ch13-\#7187 (EPIC\,211380061). Panels
  from (a) to (e) show the procedure for obtaining the
  systematic-corrected LC starting from the raw LC. Dashed grey line
  is the boundary line between the two segments in which the LC has
  been splitted during the detrending phase. In panel (f) we show cell
  and grid-point locations around the star Ch13-\#7187 loci on the
  Ch13 over the entire {\it K2}/C5. The coloured cells (size
  $0.025\times0.025$-pixel$^2$ each) represent the correction applied to
  the flux at a given $(x,y)$ position. The black-square region is
  zoomed in panel (g): for a $(x,y)$ position at a given time (magenta
  asterisk), the correction is computed by using a bi-linear
  interpolation of the four surrounding grid points (see text for
  details). We excluded the points associated to thruster-jet from the
  LCs plotted in panels (a)-(e). \label{fig:2}}
\end{figure*}

Given a raw LC normalised to its median flux, as that shown in panel
(a) of Fig.~\ref{fig:2} (star \#\,7187 in the input catalogue of Ch13,
EPIC\,211380061), and the CBV$_{i}$, with $i=1,...,16$, our routine finds
the coefficients $A_{i}$ that minimise the expression:
\begin{equation}
F_{\rm raw}^j-\sum_i\left(A_i \cdot {\rm CBV}_i^j \right)
\end{equation}
where $F_{\rm raw}^j$ is the raw flux at time $j$, $j=1,...,N_{\rm
  epochs}$, and $N_{\rm epochs}$ is the number of points in the LC. For
the minimisation, we used the Levenberg-Marquardt method
(\citealt{MINPACK-1}). In panel (b) of Fig.~\ref{fig:2} we show the
cotrended LC. It is clear that most of the systematic effects
are corrected. 

After cotrending the LCs, we detrend them for residuals systematic
errors using the same procedure as described in detail in
\citetalias{2016MNRAS.456.1137L}.  This method consists in a 2D
self flat-fielding, similar to existing techniques developed by others,
but that takes advantage of our high-precision positioning, a result of
our careful ePSF modelling and of the local-transformations approach
between the input list and the single-image catalogues.

For each target star, we modelled its median-normalised-flux LC in order
to disentangle the true intrinsic stellar variability from the
systematic errors above described.
In order to obtain the model, we divided the LC in $N-1$ segments (where
$N$ is the number of thruster firings during the entire campaign). Each
segment contained the photometric points collected between two
consecutive spacecraft thruster firings. We have identified the
``break-points'' between two segments thanks to the variations of the
target positions $\left(X,Y\right)$. In each segment we calculated the
$3.5\sigma$-clipped average of the photometric points, obtaining $N$
knots. We obtained the final model of the intrinsic variability by a
linear interpolation of the knots over the observing time (panel c of
Fig.~\ref{fig:2}).

After correcting for the intrinsic variability, the model-subtracted
LC reflected the systematic effect originated by the motion of the
star on the detector\footnote{We want to emphasise that this effect was
  in part already corrected during the contrending-phase.}  (panel d
of Fig.~\ref{fig:2}).
This is corrected according with \citetalias{2016MNRAS.456.1137L}
recipes.  Briefly, we divided the pixels ``touched'' by the target
star into an array of $40\times 40$ cells. We filled the grid by
computing the $3.5\sigma$-clipped median of the LC flux in each
element of the grid (panel f of Fig.~\ref{fig:2}).
For each $\left(x,y\right)$ position on the CCD, the correction is
given by the bi-linear interpolation between the 4 closest grid
points, as shown in panel (g) of Fig.~\ref{fig:2}.  After different
tests, we found that for M\,67 {\it K2/C5} LCs the best detrending was
achieved by splitting the time-series in two distinct segments (the
boundary between the two LC segments is marked by a dashed grey lines
in panels a-e of Fig.~\ref{fig:2}, corresponding to $\sim 37.2$\,days
after the beginning of the campaign).  Our detrending is an iterative
procedure, in such a way that both the model for the intrinsic
variability and the spacecraft drift are improved at each step. The
corrected LC is shown in panel (e) of Fig.~\ref{fig:2}.
This correction is far from being perfect and it could be considered only 
preliminary. For example, the cotrend stage works well
for a large sample of LCs, but there are variable stars  for which
the use of all the 16 CBVs is not the best solution. Indeed, the best
solution is an ad-hoc combination of CBVs for each LC that both
preserves the intrinsic stellar variability and gives the higher
photometric precision. Since we checked that a large part of
stellar LCs in our sample preserves their intrinsic signals, we
postpone the development of new LC-correction techniques to future
works. We release the raw LCs to the community to stimulate the
development of independent detrending algorithms that could be tested in the
meantime.

\begin{figure*}
\includegraphics[bb= 15 313 547 708, width=0.9\textwidth]{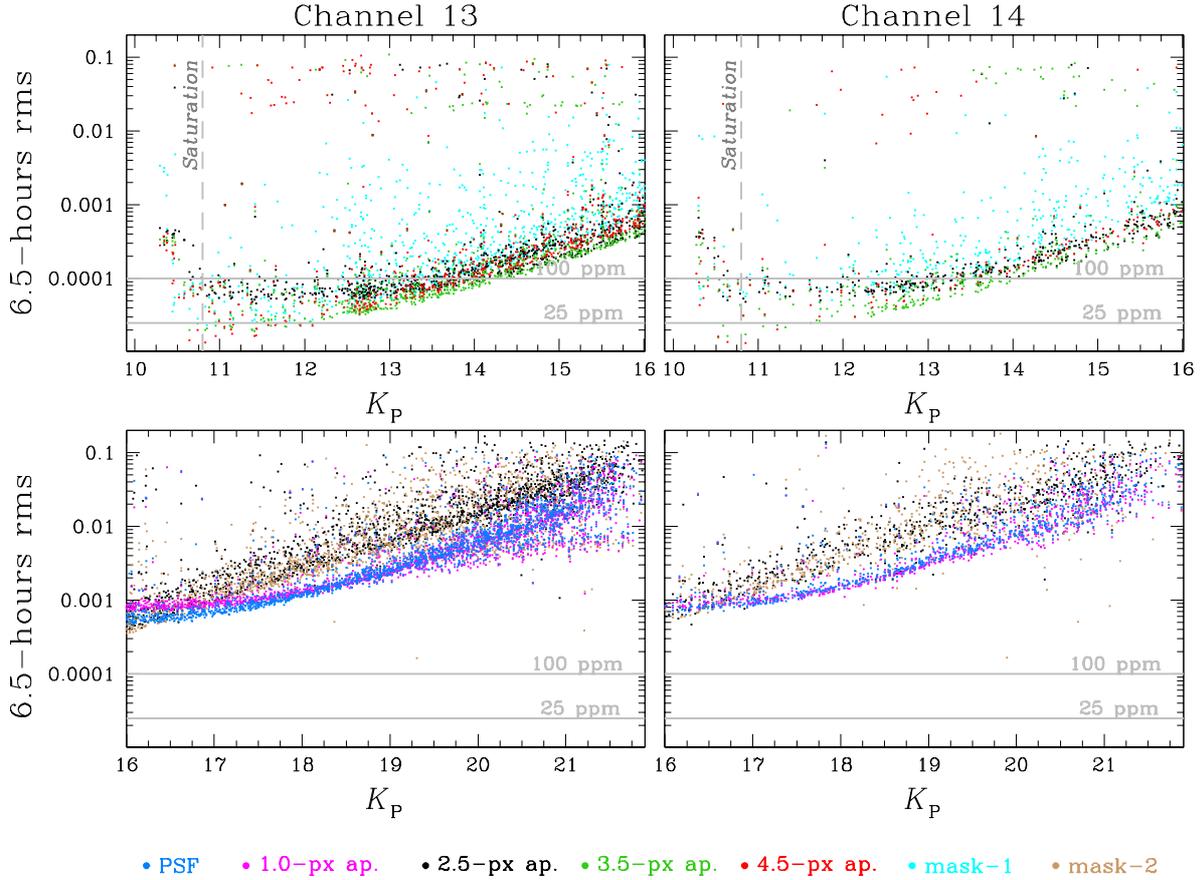}
\caption{ The 6.5h-rms for PSF-fitting (azure), 1-pixel-aperture
  (magenta), 2.5-pixel-aperture (black), 3.5-pixel-aperture (green),
  4.5-pixel-aperture (red), mask-1 (cyan), and mask-2 (brown)
  photometry on the neighbour-subtracted LCs. Top-panels show the
  6.5h-rms for stars with $K_{\rm P} < 16$ and for 2.5-pixel-aperture
  (black), 3.5-pixel-aperture (green), 4.5-pixel-aperture (red), and
  mask-1 (cyan) photometry; bottom panels for $K_{\rm P} \ge 16$ and
  PSF-fitting (azure), 1-pixel-aperture (magenta), 2.5-pixel-aperture
  (black), and mask-2 (brown) photometry.  Stars of Ch13 and Ch14 are
  plotted in left and right panels, respectively. The grey, solid
  horizontal lines are located at 100 and 25 ppm, while vertical
  dashed lines indicate the saturation limit ( $K_{\rm P} \sim 10.8$
  ).
 \label{fig:3}}
\end{figure*}

\begin{figure*}
\includegraphics[bb= 23 220 558 708,
  width=0.9\textwidth]{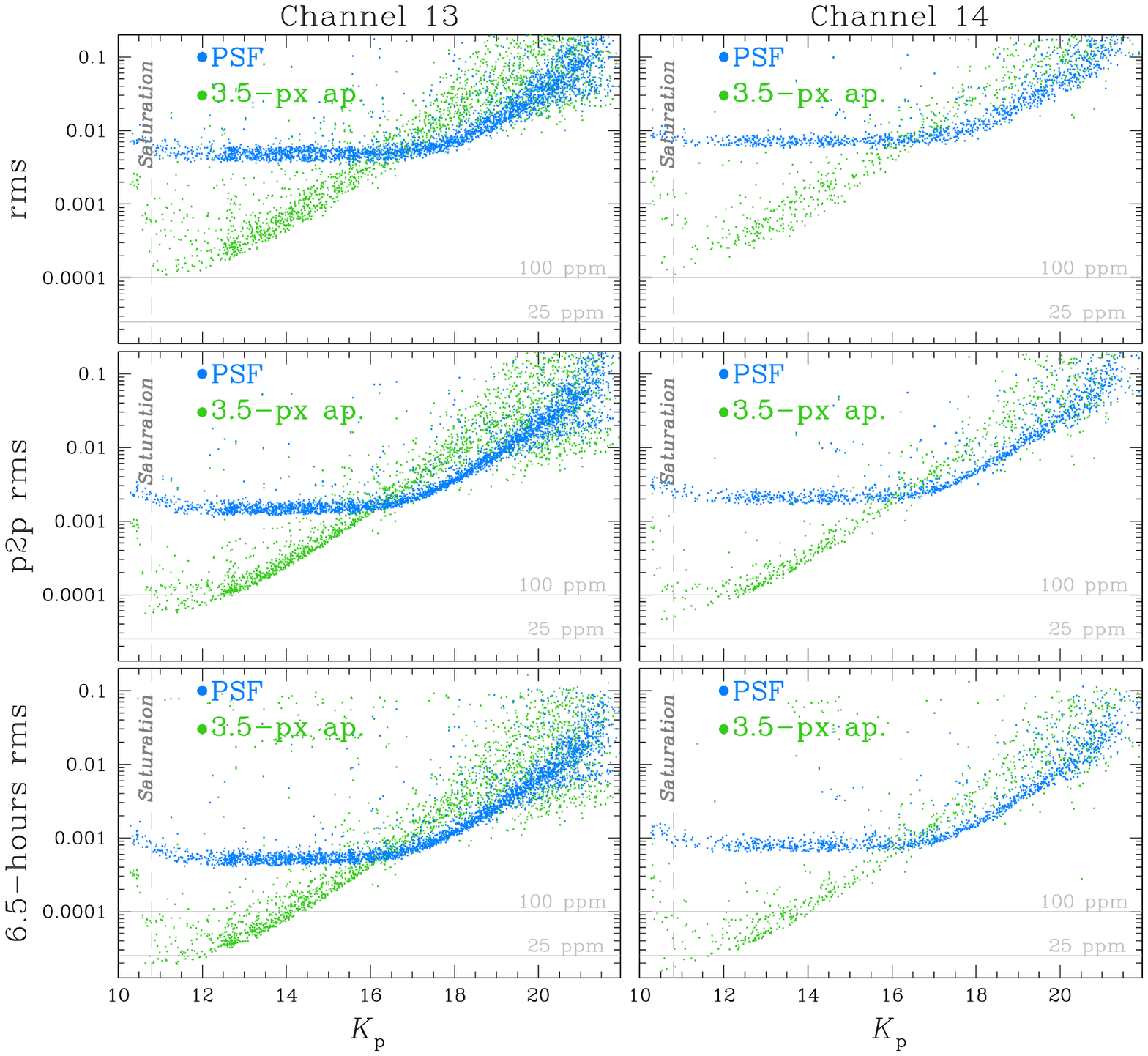}
\caption{ Photometric-precisions comparison between PSF-fitting
  (azure) and 3.5-pixel-aperture (green) photometry, that are, on
  average, the best solution in the faint- and bright-magnitude regime,
  respectively. From  top to  bottom: simple rms, p2p rms,
  and 6.5h-rms. Left and right panels show the rms for Ch13 and Ch14
  stars, respectively. Solid, horizontal lines are set at 25 and 100 ppm. Vertical
  dashed lines indicate the saturation threshold. \label{fig:3b}}
\end{figure*}

\subsection{Photometric Calibration}
We calibrated our catalogues into {\it Kepler} Magnitude System
($K_{\rm P}$) by comparing the average PSF-fitting instrumental
magnitudes of unsaturated stars with the $K_{\rm P}$-magnitudes of the
same stars in the Ecliptic Plane Input
Catalog\footnote{https://archive.stsci.edu/k2/epic/search.php} (EPIC).
We used the EPIC $K_{\rm P}$ magnitudes obtained from $gri$ photometry, as
done in Section~5 of \citetalias{2016MNRAS.456.1137L}.  We found a median difference in
zero-point of 25.31$\pm$0.07 for Ch13 and 25.17$\pm$0.06 for Ch14.

\subsection{Photometric precision}

As in \citetalias{2016MNRAS.456.1137L}, we extracted three different parameters to analyse the photometric precision:
\begin{enumerate}
\item {\it rms}: we have defined this quantity as the
  $3.5\sigma$-clipped 68.27th-percentile of the distribution around the
  median value of the points in the LC;
\item {\it point-to-point (p2p) rms}: for each LC, we have computed
  the quantity $\delta F_j = \lvert F_j-F_{j+1} \rvert$, with $F_j$ and $F_{j+1}$
  the flux values at times $j$ and $j+1$, with
  $j=1,...,N_{epochs}-1$. We have defined the p2p rms as the
  $3.5\sigma$-clipped 68.27th-percentile of the distribution around the
  median value of $\delta F$. 
\item {\it 6.5-hour rms}: we applied to each LC a 6.5h-running average
  filter. We divided the processed LC in bins containing 13
  points. For each bin, we computed the $3.5\sigma$-clipped rms and
  divided it by $\sqrt{12}$. We have defined the 6.5h rms as the median
  value of these rms measurements.
\end{enumerate}

In Fig.~\ref{fig:3} we show a comparison between the 6.5h-rms of the
different adopted photometric methods for bright (top panels) and
faint (bottom panels) stars, and for stars in Ch13 (left panels) and
Ch14 (right panels). On average, mask-1 gives the best LCs for
saturated stars ($K_{\rm P} \lesssim 10.8$), even if lower rms are
associated to 4-pixel aperture photometry. Bright, unsaturated stars
($10.8 \lesssim K_{\rm P}\lesssim 12.5$) are well measured with the
4-pixel aperture photometry (lowest 6.5h-rms $\sim$ 13.5\,ppm), while
2.5- and 3.5-pixel aperture photometric methods are the best solution
for stars with $12.5 \lesssim K_{\rm P}\lesssim 16$. In the faint
regime of magnitude ($K_{\rm P}\gtrsim 16$) 1-pixel aperture, mask-2
and PSF-fitting photometric methods give the best photometric
precision.

In Fig.~\ref{fig:3b} we show the simple rms, the p2p rms, and the
6.5h-rms for 3.5-pixel aperture and PSF-fitting photometric methods, that
are, on average, the best solution for bright and faint stars,
respectively.

\begin{figure*}
\includegraphics[bb= 15 430 569 690, width=0.9\textwidth]{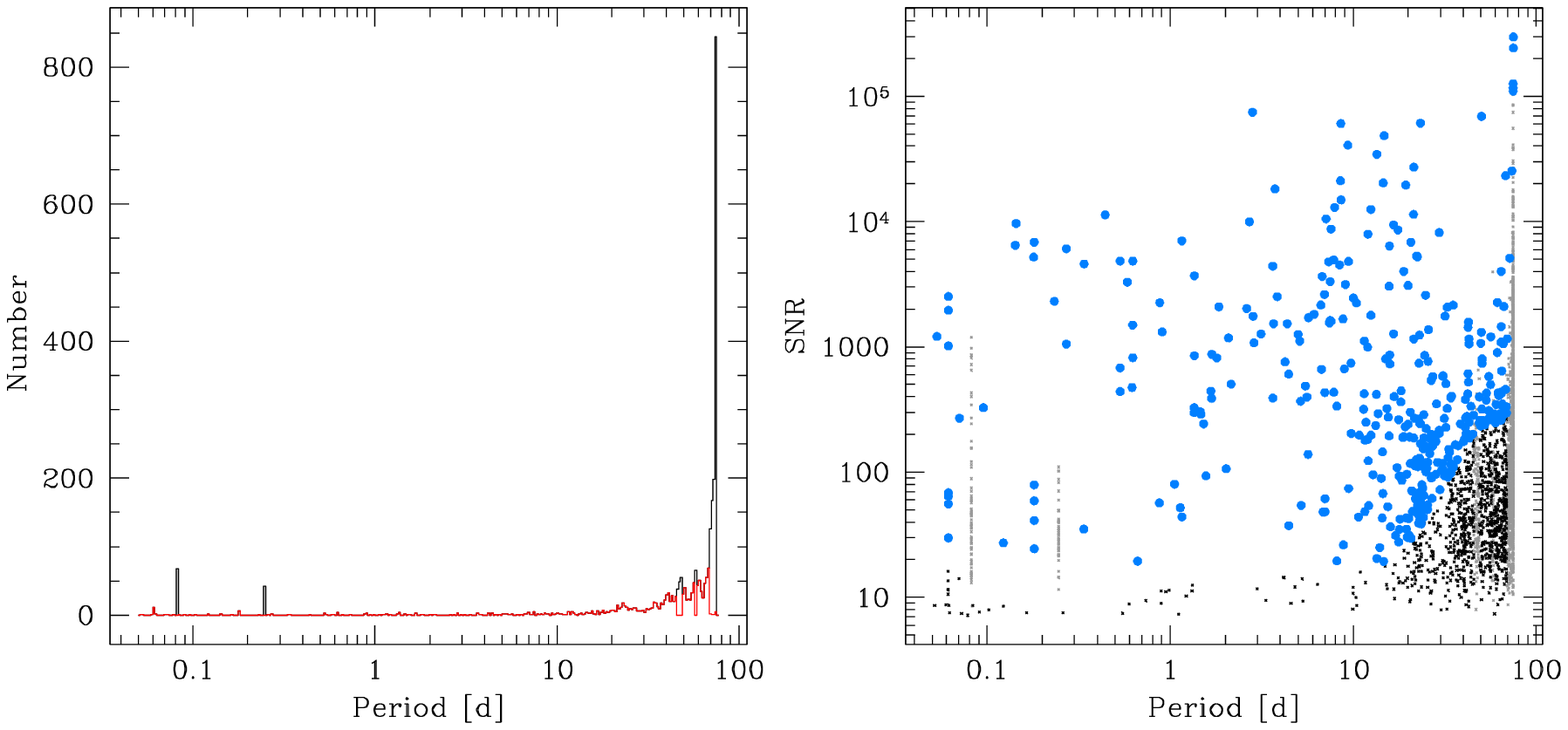}
\caption{Example of candidate-variable selection using the AoV
  algorithm (for stars in Ch13). {\it Left Panel:} distribution of the
  periods before (black) and after (red) spikes suppression. {\it
    Right Panel:} the AoV SNR as a function of the period before
  (grey) and after the spike suppression (black). In azure the
  variable candidates. \label{fig:4}}
\end{figure*}

\begin{figure*}
\includegraphics[width=0.95\textwidth]{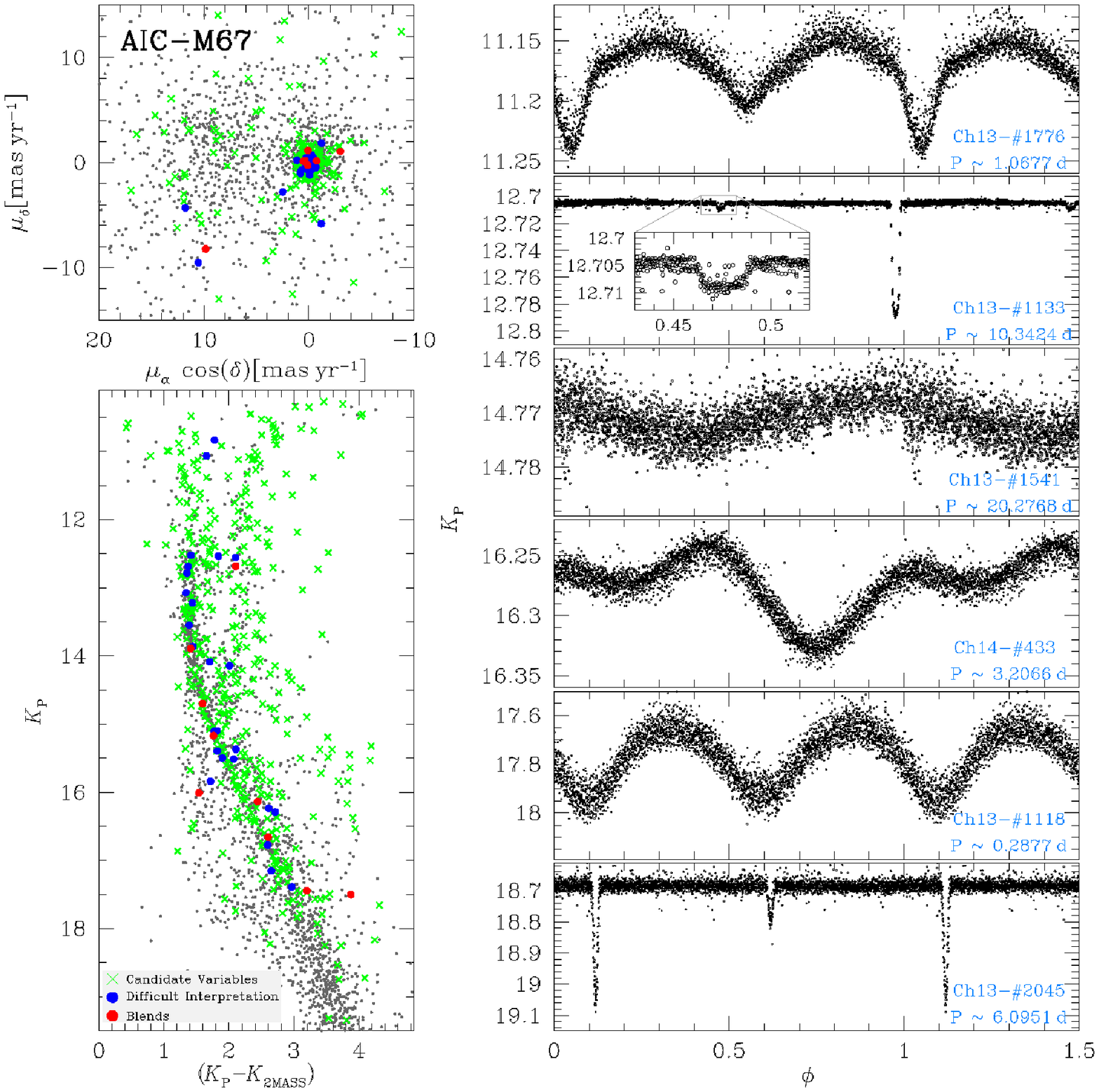}
\caption{Overview of the variability-finding results. {\it Bottom-left
    panel:} $K_{\rm P}$ versus $(K_{\rm P}-K_{\rm 2MASS})$ CMD of
  M\,67; the red dots are the blends, the green crosses the candidate
  variables, and the blue dots the stars difficult to
  interprete. {\it Top-left panel:} Vector-point diagram of proper motions (from
  \citealt{2016MNRAS.455.2337N}) for the same stars plotted in the CMD,
  colour-coded as in bottom-left panel. {\it Right panels:} a few
  examples of variable stars in order of decreasing magnitudes from 
  top to  bottom.  \label{fig:5}}
\end{figure*}

\section{Variable stars}

\label{variable}

Variable-star detection has been performed using the method
described by \citet{2015MNRAS.447.3536N,2016MNRAS.455.2337N} and also
used in \citetalias{2016MNRAS.456.1137L} and  \citet[ hereafter \citetalias{2016MNRAS.tmp.1056L}]{2016MNRAS.tmp.1056L}.

First, we cleaned the LCs from the bad points due to bad-pixels,
cosmic-rays, etc, dividing the LC in bins of 0.2~days, computing the
LC median and $\sigma$ values in each bin and clipping the points that
are 3.5$\sigma$ brighter or 15$\sigma$ fainter than the median
value. In this way we clipped-out a large part of the outliers, but
preserved the eclipsing/transits of eclipsing binaries and/or
planets. We also excluded all the thruster-jet-related events from the
LC, as done in \citetalias{2016MNRAS.456.1137L}.

We used three different algorithms (that are part of \texttt{VARTOOLS}
v.~1.33\footnote{http://www.astro.princeton.edu/$\sim$jhartman/vartools.html},
\citealt{2016A&C....17....1H}) on the clean LCs in order to detect
variable stars: the Generalised Lomb-Scargle (GLS) periodogram
(\citealt{2009A&A...496..577Z}), the Analysis of Variance (AoV)
periodogram (\citealt{1989MNRAS.241..153S}), and the Box-fitting
Least-Squares (BLS) periodogram (\citealt{2002A&A...391..369K}). Using
all these tools it is possible to detect all kind of variable stars
(sinusoidal, irregular, eclipsing binaries, etc.).  Figure~\ref{fig:4}
summarises the procedure used to isolate candidate variable stars
using the AoV method. The procedure is the same in the case of GLS and
BLS. Briefly, from the histograms of the detected periods for all the
LCs, we removed the spikes associated to spurious periods due to
systematic effects. Left panel of Fig.~\ref{fig:4} shows the histogram
before (black) and after (red) the spike suppression. Right panel of
Fig.~\ref{fig:4} shows the AoV Signal-to-noise ratio (SNR) as a
function of the detected periods, in grey and in black before and
after the spikes suppression, respectively. In this plot we selected
by hand the stars having high SNR. These points refer to stars with
high probability to be variable (azure points).  We performed the same
analysis for GLS and BLS method outputs. In the case of BLS
periodograms we used as diagnostic the Signal-to-Pink noise.  Finally,
we visually inspected each of them in order to obtain the final
catalogue.  From this catalogue, we excluded all the obvious blends by
comparing the shape and the period of each candidate-variable LC with
that of its neighbours (within 20 {\it K2} pixels).

Among a total of 4142 stars (Ch13 and Ch14) for which we have
extracted a reliable LC, we have found 318 and 170 candidate
variables for Ch13 and Ch14, respectively, for a total of 488
candidate variables. Among these candidates, we have flagged 11 stars
as obvious blends and 26 stars as difficult to interprete.  In the
difficult-interpretation sample there are sources that could be real
variable stars, blends or stars with residual systematic effects that
mime a fake variability.

In left panels of Fig.~\ref{fig:5} we show the $K_{\rm P}$ versus
$(K_{\rm P}-K_{\rm 2MASS})$ colour-magnitude diagram (CMD) of all
stars in the field (bottom) and the vector-points diagram of stellar
proper motions (top panel, from \citealt{2016MNRAS.455.2337N}
catalogue). The green crosses identify the candidate variables, the
blue dots the difficult-interpretation objects and the red dots the obvious
blends.

Finally we have cross-matched our catalogue of candidate variable
stars with the available catalogues in literature. We considered only
the 451 surely not blended stars, and we find that 152 of them have
already been catalogued by \citet{1991AJ....101..541G},
\citet{2002A&A...382..899S}, \citet{2002A&A...382..888V},
\citet{2003AJ....125.2173S,2003AJ....126.2954S},
\citet{2006MNRAS.373.1141S}, \citet{2007MNRAS.377..584S},
\citet{2007MNRAS.378.1371B}, \citet{2008MNRAS.391..343P},
\citet{2009A&A...503..165Y}, \citet{2016MNRAS.455.2337N}, and
\citet{2016MNRAS.tmp..494G}. Therefore, in our catalogue there are 299
new variable stars. Examples of variables in our catalogue are given
in right panels of Fig.~\ref{fig:5}.

\section{Candidate Exoplanet transits}
\label{candid}

To search for candidate-exoplanet transits, we used the procedure
described in detail in \citetalias{2016MNRAS.tmp.1056L}. In the following
we give a short description of our pipeline.

For each star, we flattened and cleaned its LC by modelling the
stellar intrinsic variability with a $k^{\rm th}$-order spline with
$N$ break points, removing out the outliers as described in the
previous section. To take into account different kind of variability,
we performed the analysis using three different combinations of $k$
and $N$: $k=3$ and $N=75$, $k=3$ and $N=150$, and $k=5$ and $N=175$.

For each flat/cleaned LC we extracted the BLS periodogram and we
normalised it as in \citet{2016ApJS..222...14V}, in order to
minimise the long-period false detection. We selected the five most
significant peaks in the normalised BLS periodogram, excluding the
harmonics of each peak and the spurious signals related to the
instrumentation.

For each of the five periods found, we used BLS again to refine the
central time and the duration of the candidate transit. We phased the
flat LC and checked if the transit flux drop was at least one $\sigma$
below the out-of-transit level.  Then, we verified whether there were
or not other similar flux drops in the phased LC (e.g., due to an EB).
Finally, we visually inspected the candidate transits that passed the
previous checks to exclude false alarms.

Excluding the obvious, already catalogued EBs and false alarms, we
found 5 interesting objects: 2 are EBs in the M\,67 field that,
without a proper analysis, could be mistaken for transiting-planet  host; the other 3 are candidate exoplanets. We present
them in the next sections.

\subsection{Eclipsing binaries}
\subsubsection{Star Ch13-\#1679}

\begin{figure*}
\includegraphics[bb = 18 375 592 718, width=0.95\textwidth]{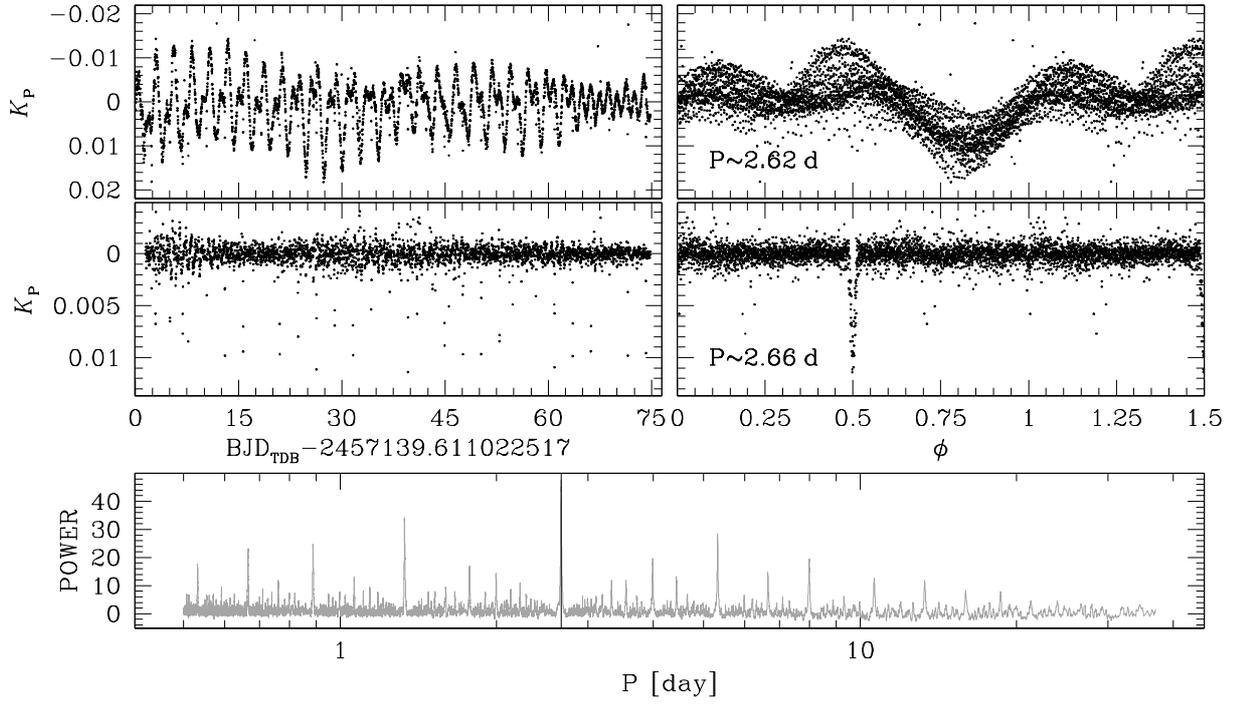}
\caption{Overview of Ch13-\#1679 LC. Top- and middle-left panels show the
  flattened LCs before and after the flattening and cleaning
  procedure, respectively. Top- and middle-right panels show the same
  LCs phased with the periods found with the AoV and BLS periodograms,
  respectively.  Bottom panel shows the BLS normalised periodogram
  (see text for details). \label{fig:6}}
\end{figure*}

The first object of interest is star Ch13-\#1679 (EPIC\,211415154,
also known as HX\,Cnc or S1070, \citealt{1977A&AS...27...89S}).
Figure~\ref{fig:6} shows an overview of its LC: top-left panel shows
the flattened LC, while the top-right panel reproduces the LC phased
with the period found by the AoV periodogram (P$\sim$2.62 d).  This is
the period related to the activity of the principal component.  We
flattened and cleaned the LC using a $5^{\rm th}$-order spline with
175 break points, and clipped out the outliers. The flattened-cleaned
LC is shown in the middle-left panel, while in the middle-right panel
the phased LC is plotted with the period (P$\sim$2.66 d) found by our
pipeline.  A careful analysis of the phased LC reveals that, in
addition to the minimum $\sim 0.013$\,mag deep, there is another
minimum $\sim 0.002$\,mag deep. The identification of this star as
an EB is reinforced by its location on the CMD of M\,67: this star,
member of M\,67 (AIC-M67 membership probability of 99.13\%), is on the sequence
of binaries. Moreover, it was classified by
\citet{2015AJ....150...97G} as double-lined binary and by
\citet{2004A&A...418..509V} as X-ray source CX48.

\subsubsection{Star Ch14-\#7224}
\label{ch147224}

\begin{figure*}
\includegraphics[bb = 18 375 592 718,width=0.95\textwidth]{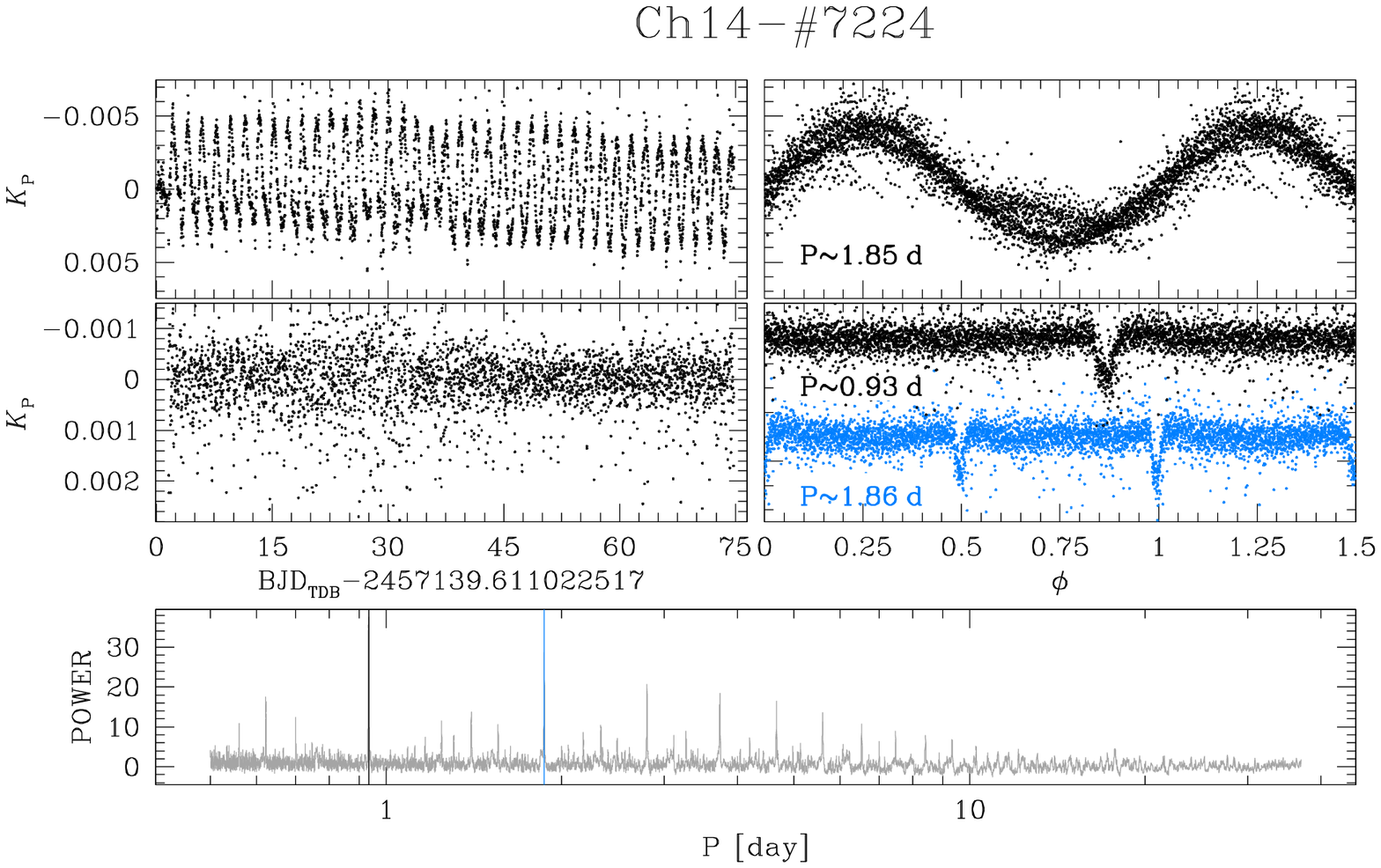}
\caption{As in Fig.~\ref{fig:6}, but for star Ch14-\#7224 \label{fig:7}.}
\end{figure*}

The second object of interest found by our pipeline is star
Ch14-\#7224 (EPIC\,211432103, Fig.~\ref{fig:7}). Indeed, in {\it K2} this source is
a blend of two stars separated by 2.70\,arcsec
(\citealt{1990ApJS...74..275H}). It belongs to the
extended input catalogue K2S-Ch14 (see Sect.\ref{aic}), derived from the
{\it K2} stacked image of Ch14, but we were not able to identify the two
components.

While the automatic classification returned a period of
P$\sim$0.93\,d, a subsequent analysis revealed its nature as EB with
period P$\sim$1.86\,d, as shown by the different minima of the azure
phased LC in the middle-right panel of Fig.~\ref{fig:7}. Again, the
period of the EB is close to that of the activity of the principal
component (P$\sim$1.85).

\begin{figure*}
\includegraphics[width=0.95\textwidth]{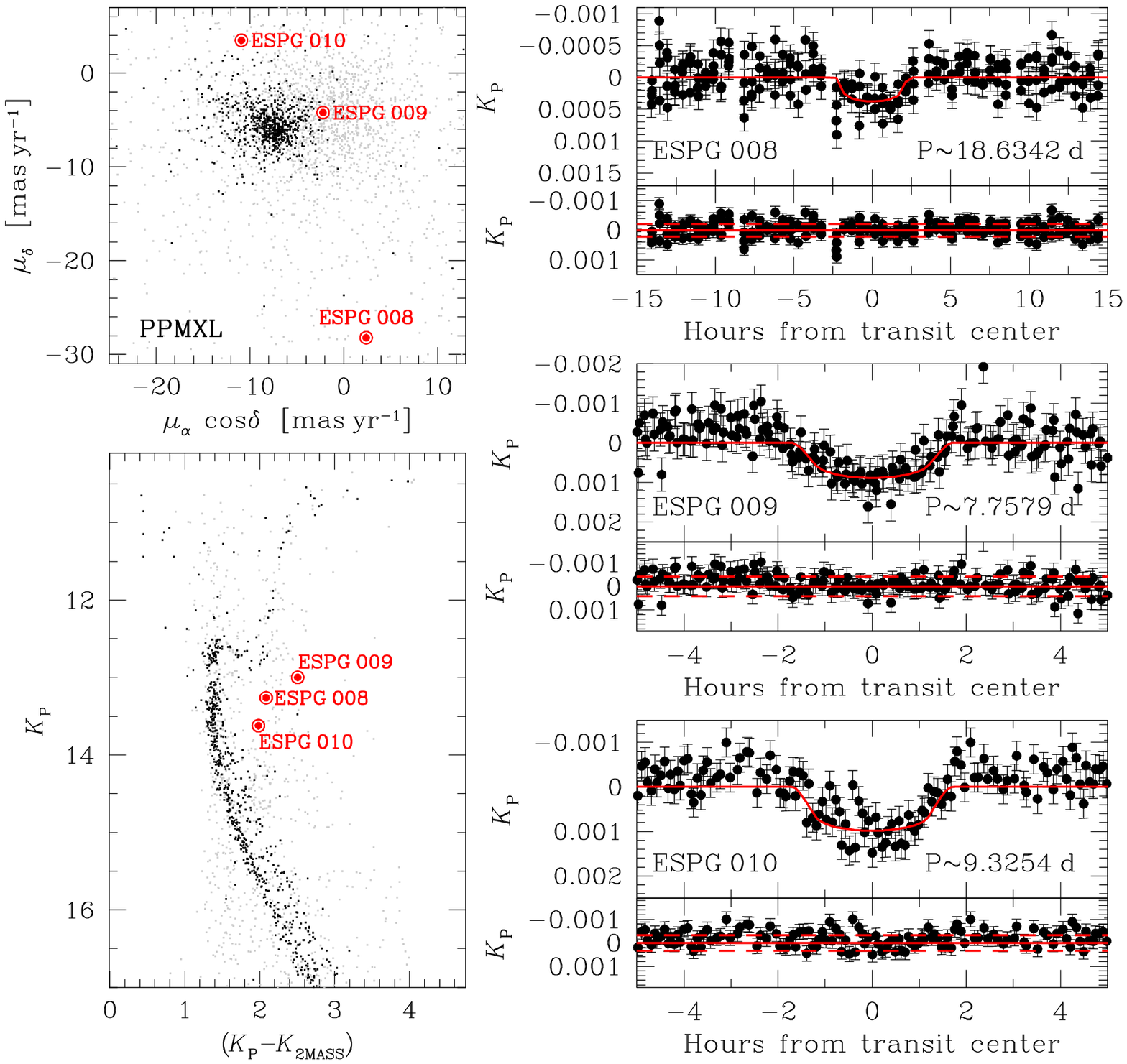}
\caption{Overview of the three candidate-exoplanets found in this
  work. {\it Bottom-left panel}: $K_{\rm P}$ versus $(K_{\rm P}-K_{\rm
    2MASS})$ CMD of M\,67; red points mark the location of the three
  candidate-exoplanet hosts. {\it Top-left panel}: PPMXL proper
  motions for the same stars shown in the bottom-left panel. In both
   panels, black points are the stars that have a membership
  probability $\ge 90$\% in the \citet{2016MNRAS.455.2337N}, in grey
  the stars that have a lower membership probability in the same
  catalogue. {\it Right panels}: the LCs of the three
  candidate-exoplanets. For each candidate, on the top, we plot the
  phased flattened LC (black dots) and the model (red line). The error
  associated to each point is the 68.27$^{\rm th}$ percentile of the
  distribution of the residual from the median value of the LC,
  excluding transit-points. On the bottom we plot the difference
  between the observed points and the model. Red solid line is the
  median of this difference, while dashed red lines correspond to $\pm
  1 \sigma$. \label{fig:8}}
\end{figure*}

\subsection{Candidate Exoplanets}

We found three candidate transiting exoplanets. Firstly, we checked
that no other variable star, with similar period, is located close
(within 100 {\it Kepler} pixels) to each candidate exoplanet.

Transit parameters were obtained using a modified version of the
particle-swarm algorithm
\texttt{Pyswarm}\footnote{https://github.com/tisimst/pyswarm} with the
\citet{2002ApJ...580L.171M} model implemented in
\texttt{PyTransit}\footnote{https://github.com/hpparvi/PyTransit}
(\citealt{2015MNRAS.450.3233P}). In order to compute the corresponding
errors, we used the
\texttt{emcee}\footnote{http://dan.iel.fm/emcee/current/} algorithm
(\citealt{2013PASP..125..306F}).
We refer the reader to \citetalias{2016MNRAS.tmp.1056L} for a detailed description of the procedure.

In our transit modelling we fixed the eccentricity $e$ to 0\,deg and
the argument of pericentre $\omega$ to 90\,deg. 
Exploiting JKTLD code (\citealt{2008MNRAS.386.1644S}), with a
quadratic law for limb darkening and the table of
\citet{2010A&A...510A..21S}, we obtained the linear and quadratic limb
darkening parameters for given $T_{\rm eff}$, $\log{g}$, and [M/H]
(see next sections), and microturbolence velocity fixed at
2\,km\,s$^{-1}$.
In the transit modelling, we
chose to derive the period ($P$), the mid-transit time of
reference ($T_0$), the inclination ($i$), and the radii ratio ($R_{\rm
  P}/R_{\rm S}$).

In the following, we give a brief description of the three
candidate-transiting exoplanets and of their parameters found by our
analysis.  In Fig.~\ref{fig:8} we show the position of the three
candidates stars on the $K_{\rm P}$ versus $(K_{\rm P}-K_{\rm 2MASS})$
CMD, their proper motions from PPMXL\footnote{The three stars belong
  to the extended input catalogues K2S-Ch13 and K2S-Ch14, and for this
  reason proper motions by \citet{2016MNRAS.455.2337N} are not
  available.}  (\citealt{2010AJ....139.2440R}), and their phased LCs.
Clearly, none of them is a M\,67 member.

In Table~\ref{tab:0} we list the parameters found of the candidate
exoplanets.  We emphasise that, as discussed in \citetalias{2016MNRAS.tmp.1056L},
the parameters found by our transit modelling are strongly dependent
on the stellar parameters adopted. Because the three stars are not
M\,67 members, we have to rely on the stellar radii and masses found
in the literature, and hence the uncertainties can be large.

\subsubsection{ESPG\,008}

Star Ch13-\#6909 (EPIC\,211439059, hereafter ESPG\,008 following
the nomenclature started in \citetalias{2016MNRAS.tmp.1056L}) is not a
 M\,67 member (as demonstrated by its high proper motion in
Fig.~\ref{fig:8}).  The depth of the transit is $\sim 310$\,ppm.

We adopted the stellar parameters listed in
{\it K2} EXOFOP website\footnote{https://exofop.ipac.caltech.edu/k2/},
 provided by \citet{2016ApJS..224....2H}.

For this star, EXOFOP gives a stellar radius $R_{\rm S} = (0.821 \pm
0.066)\,R_\odot$ and a stellar mass $M_{\rm S} = (0.877\pm 0.046)\,
M_\odot$. Using these stellar values, we found that the hosted
candidate-exoplanet has a period of $P\sim 18.6342$\,d. We found a
radii ratio $R_{\rm P}/R_{\rm S} \sim 0.0166$, corresponding to a planet radius
$R_{\rm P} \sim 1.490\,R_\oplus$.

This candidate exoplanet was already found by
\citet{2016arXiv160601264P}, and the parameters found in this paper
are in agreement with what found in their work.

\subsubsection{ESPG\,009}

The candidate exoplanet hosted by the star Ch13-\#7099
(EPIC\,211390903, S0123, hereafter ESPG\,009) is a new detection. The
star has a RV$\sim 33.03$\,km\,s$^{-1}$
(\citealt{2015AJ....150...97G}), in agreement with the mean RV of
M\,67 ($\sim 33.6$\,km\,s$^{-1}$), but from PPMXL proper motions and
the CMD, the star seems to be a field star;
\citet{1977A&AS...27...89S} found for this star a membership
probability of 38\,\%.

According to EXOFOP, this star should have a radius $R_{\rm S} =
(11.198 \pm 0.506)\,R_\odot$ and a mass $M_{\rm S} = (1.242\pm
0.132)\, M_\odot$, but using these values it is very difficult to fit
the transit. We explored the possibility that this star is not a
giant, but a K-type main sequence star. We estimated the mass and the
radius using different colour indices, the EXOFOP tabulated metallicity
(even if the final result is weakly dependent from the [M/H]) and the
empirical relations by \citet{2012ApJ...757..112B}. In this hypothesis
we found a $R_{\rm S} = (0.713 \pm 0.022)\,R_\odot$ and $M_{\rm S} =
(0.739 \pm 0.020) \,M_\odot$. Using these stellar parameters we
obtained a better model (lower $\chi^2$) than that resulted using
EXOFOP parameters. With the assumption that the star is a K-type
dwarf, we obtained that the candidate exoplanet has $P \sim
7.7579$\,d, $R_{\rm P}/R_{\rm S} \sim 0.0251$, and $R_{\rm P}\sim
1.956\,R_\oplus$.

\subsubsection{ESPG\,010}

The LC of the star Ch14-\#6981 (EPIC\,211413752, hereafter ESPG\,010)
shows candidate-exoplanet transits of depth $\sim 0.8$\,mmag.

For this star EXOFOP gives a stellar radius $R_{\rm S} = (3.666 \pm
5.924)\,R_\odot$ and a stellar mass $M_{\rm S} = (0.937\pm 0.269)\,
M_\odot$. Because the error on the stellar radius is large, we decided
to calculate the stellar radius and mass as for ESPG\,009. We found
$R_{\rm S} = (0.713 \pm 0.051) \,R_\odot$ and $M_{\rm S} = (0.738 \pm
0.046)\,M_\odot$. From our modelling, we obtained $P \sim 9.3254$\,d,
$R_{\rm P}/R_{\rm S} \sim 0.0275$, and $R_{\rm P}\sim
2.140\,R_\oplus$.

The stellar parameters adopted and the candidate-exoplanet parameters
we found are in agreement with the values obtained by
\citet{2016arXiv160601264P}.

\subsection{Summary on M\,67 exoplanets and exoplanets candidates}
\citet{2014A&A...561L...9B}, using RV measurements, discovered
planetary companions around three M\,67 members: two main sequence
stars (YBP1194 and YBP1514) and an evolved star (SAND364). Another
planet hosted by a M\,67 MS star (YBP401) was found by
\citet{2016A&A...592L...1B}.  We have verified whether these planets
are also transiting, checking their phased LCs with the period
found by \citet{2014A&A...561L...9B,2016A&A...592L...1B} and the
periods obtained with our pipeline. None of them showed transit
signature.  

\citet{2016arXiv160601264P} released a list of candidate transiting
exoplanets in {\it K2/C5} and {\it K2/C6} fields.  The two candidate
exoplanets found by \citet{2016arXiv160601264P}, located in Ch13 and
Ch14, were also found by our pipeline, namely ESPG\,008 and
ESPG\,010 above described.

Recently, \citet{2016arXiv160702339B} have released a catalogue of
candidate-exoplanets from {\it K2/C1} to {\it K2/C6}. Two of their
candidates fall in our field of view: the first candidate coincides with our
ESPG\,010; the second one is really the EB Ch14-\#7224, described in
Sect.~\ref{ch147224}.

\begin{landscape}
  \begin{table}
    \caption{Exoplanet-candidate parameters.}
    \label{tab:0}
    \begin{tabular}{ c c c c c c c c c c c c}
\hline
\multicolumn{1}{c}{ESPG} &
\multicolumn{1}{c}{EPIC} &
\multicolumn{1}{c}{$\alpha_{\rm J2000}$} &
\multicolumn{1}{c}{$\delta_{\rm J2000}$} &
\multicolumn{1}{c}{$K_{\rm P}$} &
\multicolumn{1}{c}{Period} &
\multicolumn{1}{c}{$T_0$} &
\multicolumn{1}{c}{$i$} &
\multicolumn{1}{c}{$R_{\rm P}/R_{\rm S}$} &
\multicolumn{1}{c}{$\delta_{\rm Phot}$} &
\multicolumn{1}{c}{$R_{\rm S}$} &
\multicolumn{1}{c}{$R_{\rm P}$} \\
\multicolumn{1}{c}{    } &
\multicolumn{1}{c}{    } &
\multicolumn{1}{c}{($^\circ$)} &
\multicolumn{1}{c}{($^\circ$)} &
\multicolumn{1}{c}{} &
\multicolumn{1}{c}{(d)} &
\multicolumn{1}{c}{(KBJD)} &
\multicolumn{1}{c}{($^\circ$)} &
\multicolumn{1}{c}{ } &
\multicolumn{1}{c}{(\%)} &
\multicolumn{1}{c}{($R_\odot$)} &
\multicolumn{1}{c}{($R_{\rm Jup}$)} \\
\hline

008  &  211439059   &  131.97315 & 12.232101 & 13.2615 & $18.634179 \pm 0.005229$ & $2350.798089 \pm 0.006405$ & $89.75\pm0.10$ & $0.0166 \pm 0.0005$ & $0.031 \pm 0.004$ & 0.821 & 0.133 \\
009  &  211390903   &  132.27813 & 11.498001 & 12.9985 & $07.757595 \pm 0.000822$ & $2314.806051 \pm 0.003955$ & $89.17\pm0.28$ & $0.0251 \pm 0.0007$ & $0.075 \pm 0.006$ & 0.713 & 0.174 \\
010  &  211413752   &  133.70964 & 11.848257 & 13.6226 & $09.325429 \pm 0.001094$ & $2317.180892 \pm 0.002211$ & $88.71\pm0.12$ & $0.0275 \pm 0.0007$ & $0.081 \pm 0.007$ & 0.713 & 0.191 \\

\hline\end{tabular}

    {\bf Notes:} $\delta_{\rm Phot}$ is calculated as in \citetalias{2016MNRAS.tmp.1056L}.
  \end{table}

  \begin{table}
    \caption{First six lines of Ch13 catalogue.} 
    \label{tab:1}
    \begin{tabular}{ r r r r r r r r r r r r r r }
\hline
  \multicolumn{1}{ c }{$\alpha_{\rm J2000}$} &
  \multicolumn{1}{c }{$\delta_{\rm J2000}$} &
  \multicolumn{1}{c }{$K_{\rm P}$} &
  \multicolumn{1}{c }{$B$} &
  \multicolumn{1}{c }{$V$} &
  \multicolumn{1}{c }{$R$} &
  \multicolumn{1}{c }{$I$} &
  \multicolumn{1}{c }{$J_{\rm 2MASS}$} &
  \multicolumn{1}{c }{$H_{\rm 2MASS}$} &
  \multicolumn{1}{c }{$K_{\rm 2MASS}$} &
  \multicolumn{1}{c }{ID} &
  \multicolumn{1}{c }{$\mu_{\alpha \cos \delta}$} &
  \multicolumn{1}{c }{$\mu_\delta$} &
  \multicolumn{1}{c }{P$_{\mu}$} \\
  \multicolumn{1}{ c }{($^\circ$)} &
  \multicolumn{1}{c } {($^\circ$)} &
  \multicolumn{1}{c }{   } &
  \multicolumn{1}{c }{   } &
  \multicolumn{1}{c }{   } &
  \multicolumn{1}{c }{   } &
  \multicolumn{1}{c }{   } &
  \multicolumn{1}{c }{   } &
  \multicolumn{1}{c }{   } &
  \multicolumn{1}{c }{   } &
  \multicolumn{1}{c }{   } &
  \multicolumn{1}{c }{(mas\,yr$^{-1}$)} &
  \multicolumn{1}{c }{(mas\,yr$^{-1}$)} &
  \multicolumn{1}{c }{(\%)} \\
  \multicolumn{1}{ c }{(1)} &
  \multicolumn{1}{c } {(2)} &
  \multicolumn{1}{c }{(3)   } &
  \multicolumn{1}{c }{(4)   } &
  \multicolumn{1}{c }{(5)   } &
  \multicolumn{1}{c }{(6)   } &
  \multicolumn{1}{c }{(7)   } &
  \multicolumn{1}{c }{(8)   } &
  \multicolumn{1}{c }{(9)   } &
  \multicolumn{1}{c }{(10)   } &
  \multicolumn{1}{c }{(11)   } &
  \multicolumn{1}{c }{(12)} &
  \multicolumn{1}{c }{(13)} &
  \multicolumn{1}{c }{(14)} \\
\hline
  132.96195 & 11.473325 & 18.36 & 20.71   & 19.19  & 18.32   & 17.08   & 15.61  & 14.98  & 14.81  & 0011 & 9.0118  & -17.0534 & 09.52\\
  132.74213 & 11.473428 & 17.12 & 18.89   & 17.70  & 16.97   & 16.50   & 15.11  & 14.47  & 14.28  & 0014 & -1.9178 & -2.6843  & 97.33\\
  132.73422 & 11.475366 & 15.83 & 16.99   & 16.08  & 15.64   & 15.47   & 14.29  & 13.82  & 13.76  & 0020 & -2.5016 & -2.5507  & 98.06\\
  132.73256 & 11.478913 & 15.69 & 17.53   & 16.28  & 15.61   & 14.99   & 13.58  & 12.93  & 12.81  & 0028 & 5.4254  & -18.9655 & 00.00\\
  132.49665 & 11.478087 & 13.20 & 13.76   & 13.24  & 13.07   & 13.09   & 12.10  & 11.86  & 11.79  & 0031 & 4.4767  & -2.5405  & 47.51\\
  132.54849 & 11.481701 & 17.10 & 18.79   & 17.65  & 16.92   & 16.48   & 15.10  & 14.54  & 14.40  & 0045 & -0.0164 & -4.7174  & 88.89\\
\hline\end{tabular}

  \end{table}

  \begin{table}
    \caption{First six lines of Ch13 catalogue of variables.} 
    \label{tab:2}
    \begin{tabular}{ r r r r r r r r r r r r r r }
\hline
  \multicolumn{1}{ c }{$\alpha_{\rm J2000}$} &
  \multicolumn{1}{c }{$\delta_{\rm J2000}$} &
  \multicolumn{1}{c }{$P$} &
  \multicolumn{1}{c }{$K_{\rm P}$} &
  \multicolumn{1}{c }{$B$} &
  \multicolumn{1}{c }{$V$} &
  \multicolumn{1}{c }{$R$} &
  \multicolumn{1}{c }{$I$} &
  \multicolumn{1}{c }{$J_{\rm 2MASS}$} &
  \multicolumn{1}{c }{$H_{\rm 2MASS}$} &
  \multicolumn{1}{c }{$K_{\rm 2MASS}$} &
  \multicolumn{1}{c }{ID} &
  \multicolumn{1}{c }{P$_{\mu}$} &
  \multicolumn{1}{c }{Blend} \\
  \multicolumn{1}{ c }{($^\circ$)} &
  \multicolumn{1}{c } {($^\circ$)} &
  \multicolumn{1}{c } {(d)} &
  \multicolumn{1}{c }{   } &
  \multicolumn{1}{c }{   } &
  \multicolumn{1}{c }{   } &
  \multicolumn{1}{c }{   } &
  \multicolumn{1}{c }{   } &
  \multicolumn{1}{c }{   } &
  \multicolumn{1}{c }{   } &
  \multicolumn{1}{c }{   } &
  \multicolumn{1}{c }{   } &
  \multicolumn{1}{c }{(\%)} &
  \multicolumn{1}{c }{    } \\
  \multicolumn{1}{ c }{(1)} &
  \multicolumn{1}{c } {(2)} &
  \multicolumn{1}{c }{(3)   } &
  \multicolumn{1}{c }{(4)   } &
  \multicolumn{1}{c }{(5)   } &
  \multicolumn{1}{c }{(6)   } &
  \multicolumn{1}{c }{(7)   } &
  \multicolumn{1}{c }{(8)   } &
  \multicolumn{1}{c }{(9)   } &
  \multicolumn{1}{c }{(10)   } &
  \multicolumn{1}{c }{(11)   } &
  \multicolumn{1}{c }{(12)} &
  \multicolumn{1}{c }{(13)} &
  \multicolumn{1}{c }{(14)} \\
\hline
  132.73422 & 11.475366 & 14.8451 & 15.83 & 16.99 & 16.08 & 15.64 & 15.47 & 14.29 & 13.82 & 13.76 & 0020  & 98.06 & 1\\
  132.78847 & 11.510606 & 34.4798 & 15.94 & 17.40 & 16.30 & 15.80 & 15.43 & 14.14 & 13.58 & 13.47 & 0127  & 98.91 & 1\\
  132.42671 & 11.508618 &  6.6619 & 11.36 & 11.88 & 11.38 & 11.13 & 11.47 & 10.33 & 10.10 & 10.05 & 0128  & 02.95 & 1\\
  132.72106 & 11.522206 &  5.7137 & 13.36 & 14.03 & 13.44 & 13.24 & 13.22 & 12.25 & 12.01 & 11.92 & 0164  & 99.13 & 1\\
  132.38670 & 11.530417 & 74.8200 & 11.38 & 12.00 & 11.61 & 11.12 & 11.42 & 10.14 &  9.80 &  9.72 & 0194  & 93.59 & 1\\
  132.80516 & 11.540195 & 74.8200 & 17.09 & 18.93 & 17.63 & 17.00 & 16.46 & 15.11 & 14.39 & 14.19 & 0208  & 97.60 & 1\\
\hline\end{tabular}

  \end{table}

\end{landscape}

\section{Electronic material}
\label{em}

We release\footnote{http://groups.dfa.unipd.it/ESPG/Kepler-K2.html}
raw and detrended LCs of all the sources extracted using
aperture and PSF photometric methods.

We also make public the two astro-photometric catalogues of all
sources for which we extracted the LCs, one for each analysed
channel. The catalogues contain the following information: Cols (1)
and (2) are the J2000.0 equatorial coordinates in decimal degrees;
Cols (3)-(10) are the calibrated $K_{\rm P}BVRIJ_{\rm 2MASS}H_{\rm
  2MASS}K_{\rm 2MASS}$ magnitudes (when the magnitude is not
available, it is flagged with $-$99.999); Col. (11) is the
identification number of the star; Cols (12) and (13) are the AIC-M67
relative proper motions in mas\,yr$^{-1}$ along $(\alpha \cos \delta,
\delta)$ direction (when it is not available, it is flagged with
$-$999.9999); Col. (14) is the membership probability (when it is not
available, it is flagged with $-1$). Table~\ref{tab:1} is an example
of the first six rows of the Ch13 catalogue.

We release also two catalogues (one for each analysed channel)
containing the variable stars: Cols (1) and (2) are the J2000.0
equatorial coordinates in decimal degrees; Col. (3) is the period
found (when the variability is irregular, the period is equal to
74.82); Cols (4)-(11) give the calibrated $K_{\rm P}BVRIJ_{\rm
  MASS}H_{\rm MASS}K_{\rm MASS}$ magnitudes (when the magnitude is not
available, it is flagged with $-$99.999); Col. (12) is the
identification number of the star; 
Col. (13) is the membership probability (when it is not available, it
is flagged with $-1$); Col. (14) is a flag that describe our
classification of LC: flag=1 high probability to be a real variable
star, flag=2 difficult interpretation, flag=3 high probability to be a
blend.  Table~\ref{tab:2} is an example of the first
six rows of the Ch13 catalogue of variable stars.

Finally, the publicly available electronic material contains also the
two $K2$ astrometrized stacked images (Fig.~\ref{fig:1}).

\section{Summary}
\label{sum}

In this work we presented {\it K2} LCs extracted
from images collected during the {\it K2}/C5. We have focused on a
region containing the super-stamps that cover the OC M\,67 and on all
the TPFs in Ch13 and Ch14.

For the LC extraction, we followed the same approach described in
\citet{2016MNRAS.456.1137L}, based on the use of an high-angular
resolution input catalogue, local transformations, effective
time-perturbed PSFs and on the subtraction of neighbour stars. Our
method is very efficient for extracting LCs of stars located in
crowded regions (such as M\,67 in {\it K2} images) and for faint
stars ($K_{\rm P}>15.5$).

We searched for variable stars among the 4142 extracted LCs, finding a
total of 451 variables. Of these objects, 299 are new detection.  We
found three candidate transiting exoplanets, one of them is a new
detection. All the host stars seems to be field stars rather than
M\,67 members.

We release to the community all raw and detrended LCs. This is the
first complete {\it K2} data-set of stellar LCs for the M\,67
super-stamp region. We also release the astro-photometric catalogues
of all the sources and of the identified variable stars, as well as
the astrometrized stacked images.

Our PSF-based approach is suitable for any kind of data, both ground-
and space-based observations. The work on {\it K2} data we carried out
in this series, is also a benchmark to be ready for the future space
missions focused on the search for exoplanets, such as TESS
(Transiting Exoplanet Survey Satellite, \citealt{2014SPIE.9143E..20R})
and PLATO (PLAnetary Transits and stellar Oscillations,
\citealt{2014ExA....38..249R}) .

\section*{Acknowledgements}

DN, ML, LRB, GP, LM, VN, VG, LB acknowledge PRIN-INAF 2012 partial
support by he project entitled: ``The M\,4 Core Project with Hubble
Space Telescope''. DN and GP also acknowledge partial support by the
Universit\`a degli Studi di Padova Progetto di Ateneo CPDA141214
``Towards understanding complex star formation in Galactic globular
clusters''.  ML acknowledges partial support by PRIN-INAF 2014 ``The
kaleidoscope of stellar populations in Galactic Globular Clusters with
Hubble Space Telescope''. L.M. acknowledges the financial support
provided by the European Union Seventh Framework Programme
(FP7/2007-2013) under Grant agreement number 313014 (ETAEARTH).  We
warmly thank the anonymous referee for the prompt and careful reading
of our manuscript.




\bibliographystyle{mnras}
\bibliography{biblio} 

\bsp	
\label{lastpage}
\end{document}